# NUEVAS TÉCNICAS DE TRANSMISIÓN DE MUY ALTA VELOCIDAD PARA REDES DE COMPUTADORES: Una aproximación mediante ondículas

## New technologies for high speed computer networks: a wavelet approach

Ing. José Avelino Manzano Lizcano, *Estudiante PhD-ETSIT-UPM-España,*
*Ingeniero de Telefónica de España, joseavelinomanzanolizcano@telefonica.es*
Ing. y MSc. Samuel Ángel Jaramillo Flórez, *Estudiante PhD-ETSIT-UPM-España,*
*Profesor Titular-IEE-UPB, samuel@logos.upb.edu.co*

**Abstract**

Indoor multpropagation channel is modeled by the Kaiser electromagnetic wavelet. A method for channel characterization is proposed by modeling all the reflections of indoor propagation in a kernel function instead of its impulse response. This led us to consider a fractal modulation scheme in which Kaiser wavelets substitute the traditional sinusoidal carrier.

## 1. Introducción

Cuando se habla de transmitir señales se suele referir a la transmisión de información codificada que modula a una onda electromagnética, la cual actúa de portadora y que viaja a la velocidad de la luz a través del medio que se llama canal de transmisión, siendo luego demodulada y decodificada por el receptor. Sin embargo, el hecho de que las ondas electromagnéticas estén sujetas a las leyes físicas de propagación y dispersión produce un efecto en el mensaje que a veces, cuando se intenta alcanzar los límites que impone el canal, llega a ser un "código" en si mismo, de hecho un código o mensaje absoluto que viene impuesto por el medio. Incluso en algunas aplicaciones como el radar, el medio es el propio mensaje [11], [12]. En las actuales redes de telecomunicación, la necesidad cada vez más imperiosa de alcanzar un gran ancho de banda provoca que afecte más esa limitación del canal, anterior al límite de Shannon. Por lo tanto, la búsqueda de sistemas de modulación más complejos que soslayen esa "deformación" del mensaje que produce el canal es esencial para los proveedores de servicios de telecomunicación, poseedores en muchos casos de infraestructuras antiguas (redes de pares en el caso de las compañías telefónicas) o para los fabricantes de equipos para redes de ordenadores, que buscan aprovechar al máximo el ancho de banda de radiofrecuencia que los reguladores les asignan, o que intentan desarrollar equipos que trabajen en zonas no reguladas todavía, como la banda del infrarrojo. Esto obliga a técnicas más ingeniosas para atacar el diseño de sistemas de transmisión. Ejemplos conocidos son los módems ADSL que alcanzan velocidades de transmisión por un par metálico hasta hace pocos años consideradas imposibles. Las operadoras telefónicas están muy interesadas en el avance de estas tecnologías, pues pueden así amortizar sus antiguas redes de cobre que virtualmente llegan a todos los rincones del planeta y ofrecer servicios de telecomunicación de banda ancha en consonancia con lo que demanda el mercado.

Este artículo presenta una propuesta general para el desarrollo de dichos sistemas de modulación, tomando como ejemplo de "canal a combatir" el típico canal por multipropagación en interiores que aparece cuando se intenta transmitir una señal en una habitación cerrada. Es similar al canal con multipropagación en una ciudad, donde la señal que recibe el receptor (un teléfono móvil por ejemplo) esta perturbada por múltiples reflexiones de las paredes de otros edificios, etc. Este tipo de canal es el más habitual en el caso de las transmisiones aéreas, sean por radio o por infrarrojos, y se le denomina "canal Rayleigh". Se estudia la transmisión de señales de gran ancho de banda en una habitación, sujeta a un fuerte "fading" debido a la propagación multicamino o ecos [16]. Según el enfoque que proponemos al principio de la introducción es claro que necesitamos comprender la interacción entre los dos códigos, el de la información transmitida y el de las leyes físicas que gobiernan la propagación. Inmediatamente se nos viene a la mente la idea de medir la "respuesta en frecuencia del canal" entendiendo como tal su comportamiento ante la transmisión de señales sinusoidales de un rango de frecuencias. Así, podemos definir su "ancho de banda" por ejemplo como la diferencia entre la frecuencia máxima a la cual la amplitud recibida cae la mitad y la frecuencia mínima en la que la amplitud cae también a la mitad, respecto a un valor máximo de amplitud que se obtiene para alguna frecuencia entre las dos. Pueden definirse cualesquiera otros anchos de banda, pero todos se basan en lo mismo: transmitir una señal sinusoidal, o lo que es lo mismo, respuesta del canal a una frecuencia determinada. Lo positivo de este concepto es que resulta fácilmente medible en la práctica. Un paso más es caracterizar el canal que tenemos mediante la "respuesta al impulso". Aquí se supone que ante una hipotética señal consistente en un impulso de duración infinitesimalmente pequeña, el canal responderá de cierta forma. Esta respuesta se plasma en una función que sirve como partida para el análisis matemático posterior, pues suele llevar suficiente información sobre el canal como para que podamos diseñar un sistema de comunicación. El problema es que resulta difícil medir dicha respuesta al "impulso" y lo que se suele hacer es plantear un modelo matemático del canal que, mediante simulación por ordenador, obtenga dicha respuesta al impulso. Por poner una analogía muy simplificada: la computadora simularía la emisión de un impulso de duración infinitesimal, lo reflejaría en todos los obstáculos, paredes, etc., le sumaría un ruido, y obtendría la función de respuesta al impulso buscada. Luego el diseñador trataría de encontrar un sistema de modulación, que evaluaría mediante la función obtenida. ¿Qué sistemas de modulación pueden usarse? La gran mayoría de ellos son mejoras de los sistemas de modulación de amplitud, frecuencia y fase de una "portadora" sinusoidal. Algunos utilizan las técnicas de espectro ensanchado consistentes en cambiar la frecuencia de transmisión con cada símbolo que se transmite, y a veces hasta varias veces en el intervalo que dura un símbolo, en un intento "desesperado" de hacer que el receptor esté recibiendo a una frecuencia diferente de la del eco del símbolo que fue transmitido justo un instante anterior. Pero aquí es dónde radica el núcleo de nuestro trabajo, seguimos evaluando al canal por su respuesta al impulso y seguimos caracterizándolo por su ancho de banda, conceptos que nos llevan de forma inevitable a plantearnos el problema del diseño de un sistema de transmisión en términos que un matemático diría de análisis tiempo – frecuencia. Tiempo, por la misma idea de respuesta a un impulso "infinitesimalmente pequeño", y frecuencia, por la propia definición de ancho de banda. ¿Qué puede hacer el diseñador que se enfrenta a los requisitos cada vez mayores que los fabricantes de equipos de transmisión para redes de computadores de banda ancha piden? Por poner un ejemplo, en la banda de 2.4 GHz asignada para redes de área local inalámbricas los sistemas actualmente en uso llegan a 100Mbits/sec., utilizando para ello técnicas de espectro ensanchado. En las siguientes líneas intentaremos presentar unas bases de trabajo para los diseñadores de equipos de comunicación para redes de ordenadores, de lo que denominaremos "sistemas de transmisión por ondículas (wavelets) o sistemas de modulación en "tiempo-escala", que sean también de utilidad para los responsables de las empresas que tengan que decidir la compra o evaluar nuevos productos. Los sistemas de modulación tiempo-escala los definiremos como aquellos que en lugar de tradicional análisis por frecuencia que sienta las bases de los sistemas de modulación tradicionales, utilicen el análisis tiempo-escala en las dos fases de creación de un sistema de transmisión: evaluación del canal y elección del sistema de modulación más apropiado para el mismo. Es inevitable tener que recurrir a cierto formalismo matemático para presentar los conceptos que el análisis tiempo-escala. El elemento de partida es la formulación de las leyes electromagnéticas en términos de análisis tiempo-escala. Dicha formulación fue propuesta por Kaiser [4], obteniendo de ella la significativa conclusión de que la solución de las ecuaciones de Maxwell en el dominio tiempo-escala es una **única** familia de "ondículas" ("wavelets") que permite la construcción de cualquier onda electromagnética mediante superposición de las mismas. Las ondículas o wavelets en inglés serían pues los equivalentes a nuestras tradicionales ondas sinusoidales. Al igual que el resultado conocido de que se puede representar (casi) cualquier señal como superposición de señales sinusoidales de frecuencias determinadas (lo que conocemos como espectro de un señal o transformada de Fourier de la misma), algo similar ocurre con las ondículas y, de hecho, un **número mayor** de señales van a poder ser representadas como superposición de ondículas que como superposición de sinusoides. (Ejemplos de dichas señales son las funciones fractales). Con este esquema podemos transmitir nuestro mensaje en la fase de síntesis, o sea, escogiendo los coeficientes adecuados para la generación de la onda electromagnética por superposición de ondículas. En la fase de análisis procederemos a encontrar los coeficientes que corresponden a la onda electromagnética recibida. Esto es suficiente para decodificar el mensaje que la onda transmite. Durante mucho tiempo se ha aplicado el análisis de frecuencia en procesamiento de señal, por ser la transformada de Fourier la única herramienta disponible. Con la llegada de la WFT (Windowed Fourier Transform) o transformada de Fourier localizada, mejorada por Gabor [17] se introdujo el análisis tiempo-frecuencia. Los análisis tiempo-escala y tiempo-frecuencia están relacionados con el de Fourier pero añaden más detalle, el cual aparece en los parámetros de los vectores base: la frecuencia $\omega$ en caso de la transformada de Fourier

$$e_\omega(t) = e^{2\pi i \omega t}$$

mientras que en la WFT los vectores están parametrizados por tiempo y frecuencia, y en el caso de la transformada de ondículas lo están por tiempo y escala. Sin apelar a formalismos, resulta intuitivo ver que va a ser más rica la representación del canal con dos parámetros que con uno solo (respuesta en frecuencia) como en el caso de la clásica transformada de Fourier.

**2. Dispersión de las ondículas**

Cuando la señal emitida por el transmisor incide sobre un obstáculo del canal por el que estamos intentando transmitir se produce un efecto de dispersión, que modelaremos como suma de eventos elementales de dispersión. Para definir un evento de dispersión elemental hemos de considerar tanto una onda elemental como un dispersor elemental. Nuestra onda electromagnética elemental es $\psi_z$ que genera todas las ondículas electromagnéticas al someterla a **cualquier** transformación conforme en concreto las traslaciones, las transformaciones de un sistema de referencia inercial a otro (transformaciones de Lorentz), las rotaciones y reflexiones, las dilataciones y las inversiones espacio-tiempo. Nuestro dispersor elemental lo modelamos como un espejo que ocupa una región compacta R del espacio-tiempo $\mathbf{R}^4$. La ondícula incidente se refleja con una intensidad que representamos mediante un coeficiente de reflexión $P_r$, siendo $0 \leq P_r \leq 1$. Kaiser asume [4] que la reflexión de una ondícula es otra parametrizada por el punto donde se produce la incidencia con el espejo elemental. Reconoce sin embargo que, como todos los fenómenos ondulatorios, los detalles rigurosos son difíciles y complicados y hasta el momento no se conocen con profundidad. Cierto es que la presunción de que existe un espejo *elemental* da vía libre al resto del desarrollo y a su propuesta de aplicación a las señales de radar. Dicha suposición la haremos nosotros para aplicar sus ideas al canal por reflexión difusa en interiores. Hasta el momento la reflexión se ha modelado como "difusa" o "especular". El modelo de Kaiser consiste en construir un reflector genérico mediante una distribución de densidad de reflectores elementales especulares, reconociendo que un reflector que parezca elemental a una resolución dada pueda ser "resuelto" mostrando su estructura fina en resoluciones más altas. En el presente trabajo se parte del hallazgo de Kaiser de la explotación de la invarianza de las ecuaciones de Maxwell respecto a las transformaciones afines, obteniendo mediante la extensión al dominio complejo una ondícula canónica que mantiene su forma al aplicarle cualquier transformación conforme y que permite expresar cualquier onda electromagnética como superposición de las mismas. Nuestro aporte es aplicar esa propiedad de la ondícula canónica, $\psi_z$ u "ondícula de Kaiser", al canal por dispersión en interiores (ej. redes de área local por radio en una habitación).

### 2.1. Planteamiento del modelo de propagación

El sistema de transmisión que proponemos trata de soslayar la limitación de ancho de banda que impone la transmisión en un recinto cerrado del tamaño de una habitación. El efecto de la propagación multicamino produce una reducción del ancho de banda disponible en los sistemas de modulación tradicionales. El canal se suele modelar como fading de tipo Rayleigh [1], [2], [3]. La interferencia destructiva de los ecos debidos a los diferentes trayectos de la señal produce una distribución aleatoria de la amplitud recibida. Enfocándolo desde la nueva perspectiva que el análisis tiempo-escala nos sugiere [7], tal y como comentamos en la introducción, primero evaluaremos la respuesta del canal ante nuestros vectores base (ondículas), para luego tratar de hallar el sistema de modulación y codificación más adecuado. Ya no emitimos sinusoides, por lo tanto hemos de plantear de otra forma los conceptos y técnicas conocidas de interferencia, modulaciones, anchos de banda, etc., buscando otras formas de evaluar el canal y las prestaciones del sistema de comunicaciones completo. Pero de entrada ganamos algo muy importante que la filosofía de transmisión de ondículas ortogonales nos aporta: los ecos ya no son perjudiciales [9], sino todo lo contrario, van a añadir redundancias beneficiosas para la recepción del mensaje, si utilizamos un sistema de codificación apropiado que también puede estar basado en ondículas [5]. Se produce algo parecido a lo que en terminología de Espectro Ensanchado se denomina resolver el trayecto, por lo tanto resulta más correcto definir la estrategia que utilizamos para transmitir como "resolución de ecos" en lugar de "cancelación de ecos" [9]. Sea pues $\psi(t)$ la señal real que es enviada a la antena transmisora y que genera una señal vectorial electromagnética, la ondícula $\psi_z$ transmitida por el medio. Esta señal llega por varios caminos a la antena receptora produciendo en ella una señal real f($t$). Si *a* es la atenuación del medio y *d* es el retardo de la señal por uno cualquiera de los caminos posibles, podemos escribir f($t$) = $a\, \psi\,(t-d)$, basándonos en lo expuesto en el capítulo 2, pues las

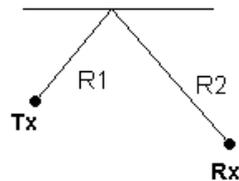

reflexiones y las transformaciones afines respetan la forma de las ondículas electromagnéticas.

siendo $d = d_1 + d_2 = R_1/c + R_2/c$. En un contexto general tanto el transmisor Tx como el receptor Rx pueden estar moviéndose a velocidades $v_1$ y $v_2$, por lo cual las distancias $R_1$ y $R_2$ son variables $R_1 = R_{10} + v_1 t$ así como $R_2 = R_{20} + v_2 t$. Como el retardo d de la señal que recibimos en el instante t es la suma de $d_1$ y $d_2$, siendo: $d_1 = 1/c * R_1(t-d)$ y análogamente $d_2 = 1/c * R_2(t-d_2)$

(ponemos $R_1(t-d)$ puesto que la señal fue emitida *d* segundos antes, y $R_2(t-d_2)$ porque la señal alcanza el reflector a los $(t-d)+d_1$ segundos, o lo que es lo mismo, a los $(t-d_2)$ segundos), con lo que podemos escribir: $d_1(t) = 1/c * R_1(t-d(t))$ y $d_2(t) = 1/c * R_2(t-d_2(t))$. De la segunda podemos despejar $d_2(t)=(R_{20}+v_2 t)/(c+v_2)$ y sustituyendo en la primera teniendo en cuenta que $d(t)=d_1(t)+d_2(t)$ tenemos

$$cd_1(t) = R_{10} + v_1 t - v_1 d_1(t) - v_1 \frac{R_{20}+v_2 t}{c+v_2}$$

con lo que tras agrupar los términos dependientes de t obtenemos

$$d = d_1(t) + d_2(t) = \frac{t[v_2(c+v_1) + v_1(c+v_2) - v_1 v_2] + R_{20}(c+v_1) + R_{10}(c+v_2) - v_1 R_{20}}{(c+v_1)(c+v_2)}$$

Podemos ahora sustituir $d$ en la expresión f($t$) = a$\psi$($t$-$d$), con lo que tras agrupar y cancelar términos llegamos a la expresión

$$f(t) = a\psi\left(\frac{t - \left(\frac{R_{10} + R_{20}}{c} + \frac{R_{20}v_2}{c^2}\right)}{\frac{(c+v_1)(c+v_2)}{c^2}}\right)$$

en la que si escogemos a = $s_0^{-1/2}$ (lo que corresponde a un factor de amplificación que el receptor puede hacer sobre el eco o señal recibidas) e identificamos

$$s_0 \equiv \frac{(c+v_1)(c+v_2)}{c^2} \qquad \tau_0 \equiv \left(\frac{R_{10} + R_{20}}{c} + \frac{R_{20}v_2}{c^2}\right)$$

obtenemos que la señal recibida es

$$f(t) = s_0^{-1/2}\psi\left(\frac{t - \tau_0}{s_0}\right)$$

## 2.2. Obtención de la distribución de reflectividad del canal

Si ahora queremos obtener la respuesta del canal, que en el plano tiempo-escala será una función de dos variables, sometamos al mismo a la propagación de la familia entera de $\psi$, $\{\psi_{s,\tau}:$ s > 0, $\tau \in \mathbf{R}\}$, de forma similar a lo propuesto en [10] con las funciones "chirp". Hagamos el producto interior de todas las $\psi_{s,\tau}(t)$ transmitidas y la f($t$) recibida

$$\tilde{f}(s,\tau) \equiv \{\psi_{s,\tau}, f\} = \int_{-\infty}^{\infty} s^{1/2} \psi\left(\frac{t-\tau}{s}\right) f(t)dt$$

que resulta ser la CWT de f(*t*) con ψ como ondícula madre. En el apartado anterior obtuvimos para la transmisión de **una única ondícula con una sola reflexión** que f(*t*) = aψ(*t-d*), siendo

$$\psi_{s,\tau}(t) \equiv s^{-1/2}\psi\left(\frac{t-\tau}{s}\right)$$

con lo que puede identificarse f(*t*) = ψ$s_0,\tau_0$(*t*). Pero en el caso de múltiples reflexiones y para todo el rango de posibles ondículas emitidas si integramos

$$f(t) = \iint \frac{ds_0 d\tau_0}{s_0^2}\psi_{s_o,\tau_0}(t)D(s_0\tau_0)$$

siendo D(s,τ) la distribución de reflectores en el plano tiempo-escala que viene determinada por el canal que estamos intentando caracterizar, y donde hemos introducido el factor $s_0^{-2}$ para simplificar desarrollos posteriores. Como ya hemos visto, podemos calcular f(s,τ) puesto que conocemos ψ y recibimos una *f* dada. La cuestión sin embargo es resolver el problema inverso, o sea hallar D(s,τ), la distribución de reflectividad del canal. Para ello, si ψ es admisible, o sea

$$C \equiv \int_{-\infty}^{\infty}\frac{d\omega}{|\omega|}|\tilde{\psi}(\omega)|^2 < \infty$$

entonces podemos recuperar *f*(*t*) a partir de su transformada CWT inversa

$$f(t) = C^{-1}\iint \frac{dsd\tau}{s^2}\psi_{s,\tau}(t)\tilde{f}(s,\tau)$$

Si comparamos esta ecuación con la expresión de *f*(*t*) obtenida anteriormente nos tienta el suponer que una posible solución para el problema inverso es

$$D(s,\tau) = C^{-1}\tilde{f}(s,\tau)$$

lo cual no va a ser cierto en general, puesto que **no toda función de reflectividad tiene porqué ser la CWT de alguna señal real f(*t*)**. Se nos plantea pues la cuestión ¿qué condiciones tiene

$$D(s,\tau) = \iint \frac{ds_0 d\tau_0}{s_0^2}K(s,\tau \mid s_0,\tau_0)D(s_0,\tau_0)$$

que tener una función D(s,τ) dada para ser D(s,τ) = g(s,τ) la CWT de alguna señal real g(*t*) función del tiempo? La función D ha de satisfacer la ecuación integral [4]

donde K es el kernel asociado a la familia de ondículas utilizada

$$K(s,\tau \mid s_0, \tau_0) \equiv C^{-1} \langle \psi_{s,\tau}, \psi_{s_0,\tau_0} \rangle$$

Esta expresión del kernel es reminiscente de la ecuación de radiosidad [8] que expresa un problema muy similar al nuestro, el de la simulación de la iluminación de objetos en entornos de realidad virtual creados por ordenador.

### 2.3. Sistema de transmisión propuesto

El esquema práctico que sugerimos para llevar a cabo nuestro sistema de transmisión consta de las siguientes etapas:

a) Fase de "resolución de ecos": se calcula la CWT del canal mediante el barrido emitiendo las ondículas $\psi s_0, \tau_0$ cubriendo sólo los extremos de escala y duración anticipados por los requisitos de velocidad de transmisión del canal.

b) Fase de "transmisión": el canal queda descrito por el campo bidimensional $D^*(s,\tau) = C^{-1} f(s,\tau)$, con lo cual podemos utilizar en principio cualquier sistema de modulación. De hecho el canal queda descrito ahora por su kernel

$$K_c(t,t') = C^{-1} \iint \frac{ds d\tau}{s_0^2} \psi_{s,\tau}(t) \tilde{f}(s,\tau) \psi_{s,\tau}(t')$$

que nos permite describirlo como un sistema lineal con respuesta

$$y(t) = S\{x(t)\} = \int_{-\infty}^{\infty} x(\tau) K_c(t,\tau) d\tau$$

Por lo tanto, hemos llegado a definir el canal mediante su respuesta al impulso, que no es otra que el kernel $K_c$ (distinto al kernel K generado por la familia de ondículas $\psi$). Pero aunque esto sea un dato importante para la caracterización de nuestro canal, no lo es tanto desde el punto de vista del desarrollo de un sistema de modulación para el mismo, pues perdemos toda la riqueza que la representación tiempo-escala nos da, siendo más conveniente describirlo como $D^*(s,\tau) = C^{-1}f(s,\tau)$ si el sistema de modulación va a poder representarse también en el plano tiempo-escala. Este es el caso de la modulación fractal propuesta en [5].

## 3. Conclusiones y futuros desarrollos

En el presente artículo, basándonos en las propiedades de las ondículas de Kaiser, proponemos su empleo como señales de transmisión, o señales portadoras, para la realización de sistemas de transmisión capaces de "resolver" la propagación multicamino soslayando así la limitación de "ancho de banda" que se hacía insalvable cuando se emplean las tradicionales sinusoides como portadoras. Sugerimos la caracterización de la respuesta del canal con multipropagación por reflexión en interiores mediante su CWT, o sea, una función de dos variables, tiempo y escala, proponiendo un método experimental para la medición de la misma, mediante un "barrido de ondículas" reminiscente del barrido de frecuencia que se hacía con un vobulador.

Quedan abiertos varios interrogantes que nos dan pie a otras tantas líneas de investigación:

- Discretización del proceso para su realización mediante procesadores digitales de señal.
- Búsqueda de sistemas de modulación para las ondículas portadoras.
- Evaluar, en este tipo de canal con multipropagación, el sistema de "modulación fractal" propuesto por Wornell [5], basado en la modulación de réplicas de la señal a transmitir sobre una base ortonormal de ondículas en todas las escalas simultáneamente.
- Obtener la CWT del canal mediante simulación por ordenador, en lugar de por métodos experimentales.
- Estudiar la idoneidad de las ondículas de Kaiser para la caracterización de estructuras fractales, que previsiblemente son las que describen la reflexión difusa que aparece en el canal, tal y como se constata cuando se trabaja en longitudes de onda del infrarrojo [16].

Proponer si se da el caso otras ondículas más apropiadas para describir medios fractales [6].

## 4. Bibliografía

[1] B. Solaiman, A. Glavieux, A. Hillion. Error Probability of Fast Frequency Hopping Spread Spectrum with BFSK Modulation un Selective Rayleigh and Selective Rician Fading Channels. IEEE Transactions on Communications. February 1990, p. 233-240.

[2] O. Yue. Frequency-Hopping, Multiple-Access, Phase-Shift-Keying System Performance in a Rayleigh Fading Environment. The Bell Technical Journal. July-August 1980, p. 861-879.

[3] M. Moeneclaey, J. Wang. Multiple hops/symbol fast frequency-hopping spread spectrum multiple access with coding for indoor radio. IEEE Proceedings-I. February 1992, p. 95-102.

[4] G. Kaiser. A Friendly Guide to Wavelets. Birkhäuser 1994.

[5] G. W. Wornell. Signal Processing with Fractals: A Wavelet-Based Approach. Prentice Hall 1996.

[6] F. J. Herrmann. A scaling medium representation, a discussion on well-logs, fractals and waves. PhD dissertation. Delft University 1997.

[7] R. E. Learned, H. Krim, B. Claus, A. S. Wilsky, W. C. Karl. Wavelet-Packek-Based Multiple Access Communication. Proceedings of the SPIE international symposium on optics, imaging an instrumentation. July 1994.

[8] P. Schröder, S. J. Gortler, M. F. Cohen, P. Hanrahan. Wavelet Projections for Radiosity. Fourth Eurographics Workshop on Rendering 1993.

[9] Y. Lu, J. M. Morris. Gabor Expansion for Adaptive Echo Cancellation. IEEE Signal Processing Magazine. March 1999.

[10] M. Bernfeld. Chirp Doppler Radar. Proceedings of the IEEE. April 1984.

[11] V. C. Chen. Radar ambiguity function, time-varying matched filter, and optimum wavelet correlator. Optical Engineering. July 1994.

[12] V. C. Chen. Aplications of time-frequency processing to radar imaging. Optical Engineering. April 1997.

[13] S. Qian, V. C. Chen. Joint Time-Frequency Transform for Radar Range-Doppler Imaging. IEEE Transactions on Aerospace and Electronic Systems. April 1998.

[14] L. Nottale. Scale relativity and fractal space-time: applications to quantum physics, cosmology and chaotic systems. Chaos, solitons and fractals, 7 (6). p. 877-938. 1996.

[15] L. Nottale. El espacio-tiempo fractal. Investigación y Ciencia. Julio 1997.

[16] A. Santamaría, F. J. López-Hernández. Wireless LAN Systems. Artech House 1994.

[17] D.Gabor. Theory of communication. J. Inst. Electr. Eng. 93 (III) p. 429-457. 1946.

[18] T. C. Bailey, T. Sapatinas, K. J. Powell, W. Krzanowski. Signal Detection in Underwater Sound Using Wavelets. Journal of the American Statistical Association. 1997.

[19] A. Davis, A. Marshak, W. Wiscombe. Wavelet-Based Multifractal Analysis of Non-Stationary and/or Intermittent Geophysicla Signals. Wavelets in Geophysics. August 1994.

[20] D. Jaggard. Prolog to Special Section on Fractals in Electrical Engineering. Proc. IEEE 81, p. 1423-1427. 1993.